\documentclass[submission, Proceedings]{SciPost}

\hypersetup{
    colorlinks,
    linkcolor={red!50!black},
    citecolor={blue!50!black},
    urlcolor={blue!80!black}
}

\usepackage[bitstream-charter]{mathdesign}
\usepackage[rightcaption]{sidecap}
\sidecaptionvpos{figure}{c}
\DeclareSymbolFont{usualmathcal}{OMS}{cmsy}{m}{n}
\DeclareSymbolFontAlphabet{\mathcal}{usualmathcal}

\begin{document}

\begin{center}{\Large \textbf{
Di-\textbf{$\pi^0$} correlations in $p+$$p$, $p+$$\mathrm{Al}$ and $p+$$\mathrm{Au}$ collisions at $\sqrt{s_{\mathrm{NN}}}$ = 200 GeV at STAR\\
}}\end{center}

\begin{center}
Xiaoxuan Chu\textsuperscript{1}, 
\\
for the STAR Collaboration
\end{center}

\begin{center}
{\bf 1} Brookhaven National Laboratory, Upton, New York 11973, USA
\\
* xchu@bnl.gov
\end{center}

\begin{center}
\today
\end{center}



\definecolor{palegray}{gray}{0.95}
\begin{center}
\colorbox{palegray}{
  \begin{tabular}{rr}
  \begin{minipage}{0.1\textwidth}
    \includegraphics[width=22mm]{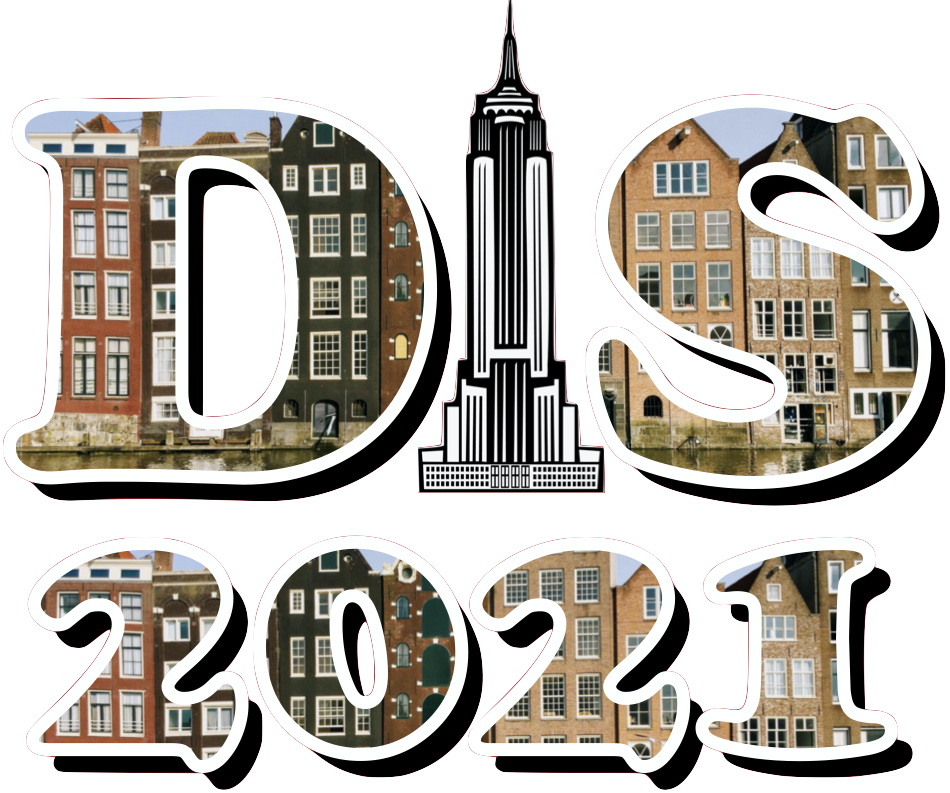}
  \end{minipage}
  &
  \begin{minipage}{0.75\textwidth}
    \begin{center}
    {\it Proceedings for the XXVIII International Workshop\\ on Deep-Inelastic Scattering and
Related Subjects,}\\
    {\it Stony Brook University, New York, USA, 12-16 April 2021} \\
    \doi{10.21468/SciPostPhysProc.?}\\
    \end{center}
  \end{minipage}
\end{tabular}
}
\end{center}

\section*{Abstract}
{\bf
The STAR Collaboration reports measurements of back-to-back azimuthal correlations of di-$\pi^0$ produced at forward pseudorapidity ($2.6<\eta<4.0$) in $p+$$p$, $p+$$\mathrm{Al}$ and $p+$$\mathrm{Au}$ collisions at a center-of-mass energy per nucleon-nucleon pair of 200 GeV.
A clear suppression of the correlated away-side yields is observed in $p+$$\mathrm{Au}$ and for the first time in $p+$$\mathrm{Al}$ collisions, compared with the $p$$+$$p$ data. The enhanced suppression found in $p+$$\mathrm{Au}$ with respect to $p+$$\mathrm{Al}$ collisions exhibits the saturation scale ($Q_s^2$) dependence on the mass number~~$A$. The observed suppression of back-to-back pairs as a function of event activity and transverse momentum points to non-linear gluon dynamics arising at high parton densities.}


\section{Introduction}
\label{sec:intro}
Collisions between hadronic systems, i.e., $p+$A and $d+$A at the Relativistic Heavy Ion Collider (RHIC) provide a window to the parton distributions of nuclei at small momentum fraction ($x$). Several RHIC measurements~\cite{BRAHMS:2003sns,Arsene:2004ux,Adler:2004eh,Adams:2006uz,Adare:2011sc} have shown that the hadron yields at forward rapidities (deuteron going direction) are suppressed in $d+$Au collisions relative to $p$+$p$ collisions at $\sqrt{s_{\mathrm{NN}}}=200$~GeV. 
~Possible mechanisms leading to the suppression include gluon saturation~\cite{Gribov:1984tu} and energy loss~\cite{Vitev:2007ve,Kang:2011bp}. 
Meanwhile, the contributions from double-parton interactions to the $d+$A$\rightarrow \pi^0 \pi^0$X cross section are suggested as an alternative explanation of the suppression~\cite{Strikman:2010bg}. 
Therefore, it is important to carry out the same measurement in $p+$A collisions, which are theoretically and experimentally cleaner compared to $d+$A collisions.

The density of
gluons per unit transverse area is expected to be larger in
nuclei than in nucleons at a given $x$; thus, nuclei provide a natural
environment to study non-linear gluon evolution. Gluons from different nucleons can interfere and amplify the total transverse gluon density by a factor of $A^{1/3}$ for a nucleus with mass number $A$.
The color glass condensate (CGC) framework~\cite{McLerran:1993ni,McLerran:1993ka} predicts that a quark or gluon scattering at forward angles (large rapidities) will interact coherently with 
gluons at low-$x$ in the nucleus~\cite{Guzey:2004zp}.  
As a result, for di-hadron correlations on the away-side, the yield of associated hadrons is expected to be suppressed and the correlation peak is predicted to be broadened in $p(d)+$A collisions compared to $p$+$p$ collisions~\cite{Kharzeev:2004bw,Marquet:2007vb}. 



In this contribution, we report measurements of back-to-back di-$\pi^0$
correlations for $p+$Al and $p+$Au collisions relative to $p$+$p$ collisions in the forward-rapidity region ($2.6<\eta<4.0$) at $\sqrt{s_{\mathrm{NN}}}=200$ GeV. The use of different ion beams provides an opportunity to study the saturation scale ($Q_s^2$) dependence on $A$. The correlation function $C(\Delta\phi) = N_\mathrm{pair}(\Delta\phi)/(N_\mathrm{trig}\times\Delta\phi)$ is measured, where $N_\mathrm{pair}$ is the yield of the correlated
trigger and associated $\pi^0$ pairs, $N_{\mathrm{trig}}$ is the
trigger $\pi^0$ yield, $\Delta\phi$ is the azimuthal angle difference between the trigger and associated $\pi^0$. In each pair, trigger $\pi^0$ is the one with higher transverse momentum $p_{T}$ ($p_{T}^{\mathrm{trig}}$); associated $\pi^0$ is the one with lower $p_{T}$ ($p_{T}^{\mathrm{asso}}$). To remove detector acceptance effects, the measured correlation functions are divided by the correlation functions from mixed events. 


\section{Experiment and Dataset}
Datasets for $p$+$p$, $p+$Al and $p+$Au collisions were recorded in 2015.
The $\pi^0$s were reconstructed from photons, which were identified with the STAR forward meson spectrometer (FMS). The FMS is an electromagnetic calorimeter covering a pseudorapidity range from 2.6 to 4.0~\cite{STAR:2018iyz}.
The collision events are triggered by FMS based on the transverse energy deposition. 
The $p+$Al and $p+$Au samples are separated into different event activity (E.A.) classes based on the 
energy ($\Sigma E_\mathrm{BBC}$) deposited in the inner sectors of the beam beam counter (BBC) at backward direction (aluminum and gold going direction, $3.3<-\eta<5.0$), where $\Sigma E_\mathrm{BBC}$ is the ADC sum from all 16 BBC tiles. The STAR BBC is a scintillator detector which measures minimum-ionizing particles~\cite{Whitten:2008zz}. The samples without any E.A. selections are minimum bias (MinBias) data. 
The energy and $p_{T}$ of the photon candidates are required to be above 1 GeV and 0.1 GeV/$c$, respectively. 
The reconstructed $\pi^0$'s $p_{T}$ is above 0.5 GeV/$c$. The energy asymmetry of $\pi^0$'s photon components $|\frac{E_{1}-E_{2}}{E_{1}+E_{2}}|$ is required to be below 0.7, where $E_{1}$ and $E_{2}$ are the photon energies. The selected mass range of the $\pi^0$ candidates is between 0.07 and 0.2 GeV/$c^{2}$. 

\section{Results}
\begin{figure}[!ht]
\vspace{-0.0cm}
\hspace{-0.5cm}
  \includegraphics[width=1.0\linewidth]{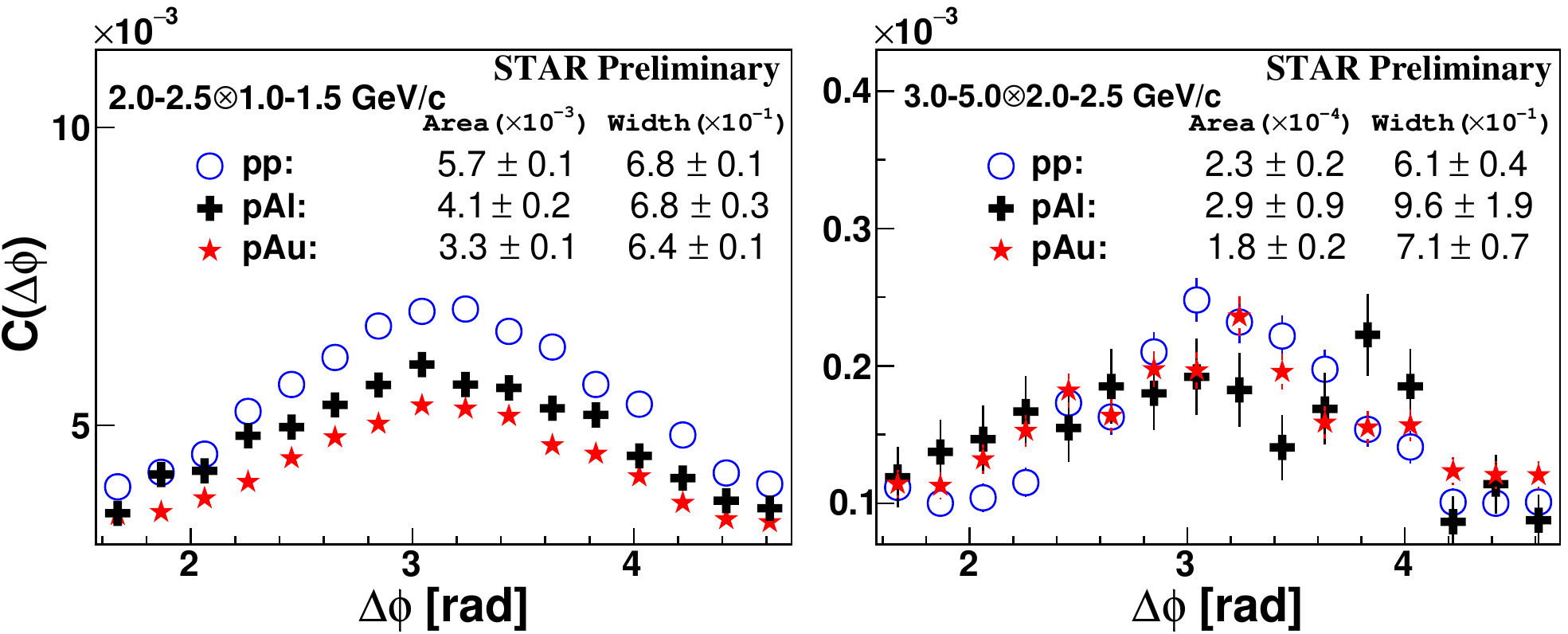}
\caption{ 
Comparison of the correlation functions 
vs. azimuthal angle difference between forward ($2.6<\eta<4.0$) $\pi^{0}$ pairs
in MinBias $p$+$p$, $p+$Al and $p+$Au collisions at $\sqrt{s_{\mathrm{NN}}}=200$ GeV.~~Left panel: the trigger $\pi^0$'s $p_{T}$ ($p^{\mathrm{trig}}_T$) = 2$-$2.5 GeV/$c$ and the associated $\pi^0$'s $p_{T}$ ($p^{\mathrm{asso}}_T$) = 1$-$1.5 GeV/$c$; right panel: $p^{\mathrm{trig}}_T$ = 3$-$5 GeV/$c$ and $p^{\mathrm{asso}}_T$ = 2$-$2.5 GeV/$c$.  The area and width of away-side peaks are shown in each panel as described in the text.
}
  \label{fig:A_corr}
\end{figure}

The corrected correlation function as described in Sec.~\ref{sec:intro} is fitted with two individual Gaussians at the near- ($\Delta\phi=$ 0) and away-side ($\Delta\phi=\pi$) peaks, together with a constant for the pedestal, from $\Delta\phi = -\pi/2$ to $\Delta\phi = 3\pi/2$. The near-side peak, dominated by two $\pi^0$s coming from the same jet, 
encodes nuclear modifications to the parton fragmentation.
This proceeding will not discuss the near-side physics, and focuses on the study of the away-side peak. The area of the away-side peak is the integral of the correlation function from $\Delta\phi = \pi/2$ to $\Delta\phi = 3\pi/2$ after pedestal subtraction, representing the back-to-back $\pi^0$ yields per trigger particle; the corresponding width is defined as the $\sigma$ of the away-side peak according to the fit.

Figure~\ref{fig:A_corr} shows the comparison of forward back-to-back di-$\pi^0$ correlation function in MinBias $p$+$p$, $p+$Al and $p+$Au collisions at $\sqrt{s_{\mathrm{NN}}}=200$ GeV. The area and width of away-side peak from different collisions are shown, together with their statistical uncertainties.
In the left panel, in the low $p_T$ regime, a clear suppression is observed in $p+$A compared to the $p$+$p$ data. The away-side associated $\pi^{0}$ yield per-trigger in $p+$Au ($p+$Al) is suppressed by about a factor 1.7 (1.2) with respect to $p$+$p$ collisions. 
The enhanced suppression in $p+$Au relative to $p+$Al at the same collision energy supports an $A$ dependence of $Q_s^2$ as predicted in \cite{McLerran:1993ni,Kharzeev:2004bw}. The suppression decreases with increasing $p_T$ of the $\pi^0$s. In the high $p_T$ range, no suppression is observed in $p+$A compared to $p$+$p$ collisions as can be seen in the right panel of Fig.~\ref{fig:A_corr}. The parton momentum fraction $x$ with respect to the nucleon inside the nucleus 
increases with the $p_{T}$ of the trigger and associated $\pi^0$s.
$Q$ can be approximated as the average $p_{T}$ of di-$\pi^0$. The low $x$ and $Q^{2}$ regime, where the gluon density is large and expected to be saturated, 
can be accessed using low $p_{T}$ di-$\pi^{0}$ pairs. When the $\pi^{0}$ $p_{T}$ is high, the probed $x$ ($Q^{2}$) will not be sufficiently small to reach a non-linear regime. 
The phenomenon of broadening is not observed in $p+$A collisions, which is consistent with the similar measurement in $d+$Au collisions by the PHENIX experiment~\cite{Adare:2011sc}.

\begin{SCfigure}
 \includegraphics[width=0.5\textwidth]{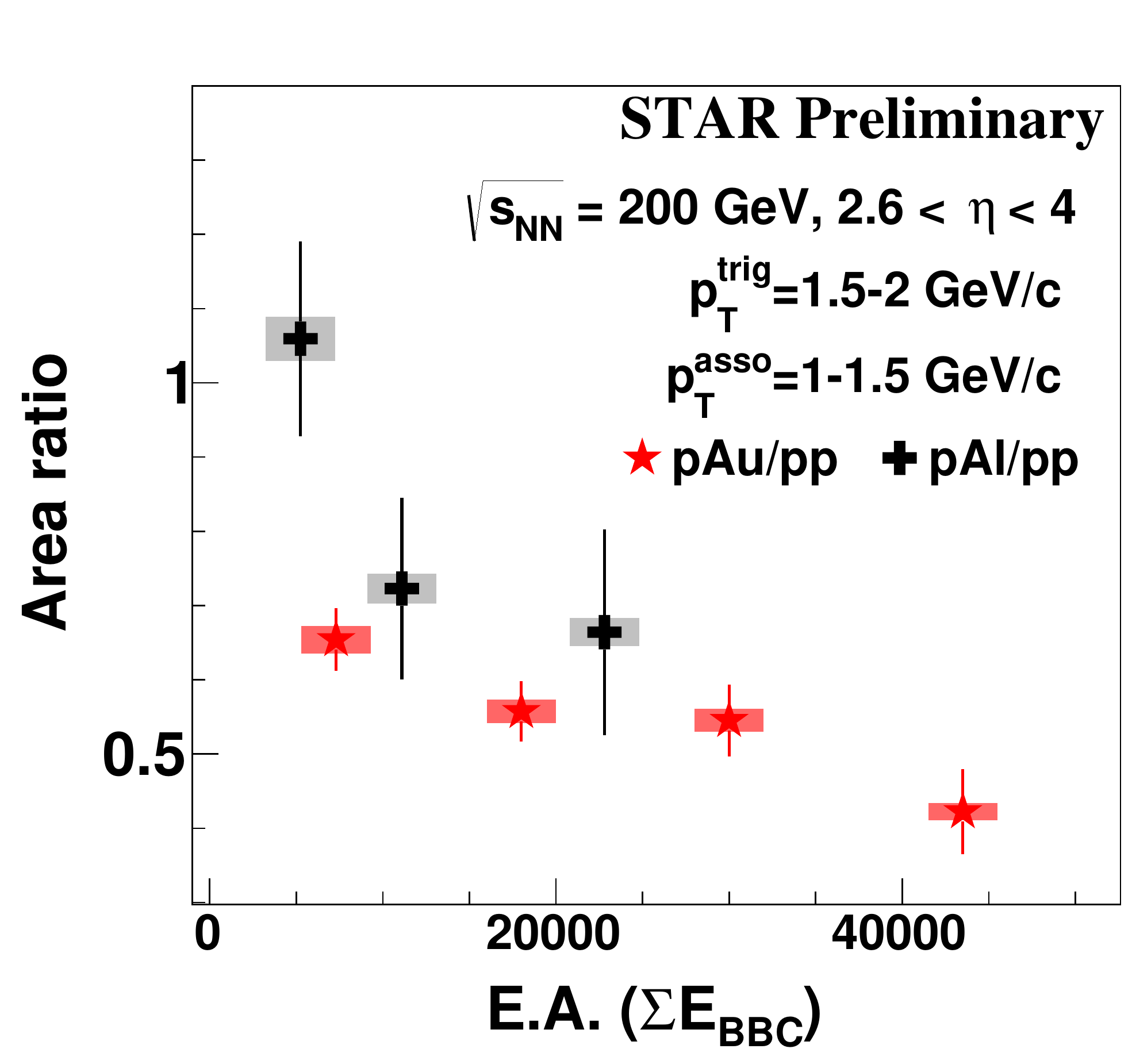}
\caption{Area ratio of away-side di-$\pi^0$ correlation at forward rapidities ($2.6<\eta<4.0$) for different event activity bins from $p+$Al and $p+$Au relative to MinBias $p$+$p$ collisions at $p^{\mathrm{trig}}_{T}$ = 1.5$-$2 GeV/$c$ and $p^{\mathrm{asso}}_{T}$ = 1$-$1.5 GeV/$c$. The vertical bars around each data point indicate statistical uncertainties and the vertical bands indicate point-to-point systematic uncertainties. The width of the band is chosen for visibility and doesn't reflect uncertainties.}
  \label{fig:sum_pAlpAu}
\end{SCfigure}

In Fig.~\ref{fig:sum_pAlpAu}, ratios of the away-side peak area for di-$\pi^{0}$ correlations from $p+$Al and $p+$Au collisions to that from MinBias p+p collisions are shown for different event activity classes.
The systematic 
uncertainties of the area arise from non-uniform detector efficiency as a function of $\phi$, and estimated as the following.
We started with a physical-like correlation without detector effects. A correlation with detector effects included was obtained by applying weights according to the $\phi$ distributions from data, and then a mixed event correction 
was applied to the correlation as done in real data analysis.
The difference between the input physical-like and the corrected correlation is taken as the systematic uncertainty.
An enhanced suppression in high activity events is observed in $p+$Au and $p+$Al data, and the significance of the stronger suppression in the highest E.A. than the lowest E.A. in $p+$Au ($p+$Al) collisions is 3.1 (1.7) $\sigma$. Less suppression is observed in $p+$Al compared to $p+$Au, which is consistent with the results at low $p_{T}$ from MinBias $p+$Al and $p+$Au data shown in the left panel of Fig.~\ref{fig:A_corr}.

\section{Conclusion}
In summary, ~measurements of azimuthal correlations of di-$\pi^0$ at forward rapidities ($2.6<\eta<$
4.0) are performed using 2015 200 GeV $p$+$p$, $p+$Al and $p+$Au data at STAR. 
A clear suppression of away-side yields is observed in $p+$A in comparison with $p$+$p$ collisions at low $p_{T}$. 
The suppression is enhanced at higher E.A. and for pairs probing smaller $x$ (and $Q^{2}$) with lower di-$\pi^0$'s $p_{T}$. 
No increase in the width of the azimuthal angular correlation is seen within experimental uncertainties. 
The presented results are the first measurement of the nuclear effect dependence on $A$, where we observe that the suppression is enhanced with larger $A$.

\section{Acknowledgements}
We thank the RHIC Operations Group and RCF at BNL. This work is supported by U.S. Department of Energy under contract number de-sc0012704.

\bibliography{ref.bib}

\begin{thebibliography}{10}
\providecommand{\url}[1]{\texttt{#1}}
\providecommand{\urlprefix}{URL }
\expandafter\ifx\csname urlstyle\endcsname\relax
  \providecommand{\doi}[1]{doi:\discretionary{}{}{}#1}\else
  \providecommand{\doi}{doi:\discretionary{}{}{}\begingroup
  \urlstyle{rm}\Url}\fi
\providecommand{\eprint}[2][]{\url{#2}}

\bibitem{BRAHMS:2003sns}
I.~Arsene \emph{et~al.},
\newblock \emph{{Transverse momentum spectra in Au+Au and d+Au collisions at
  $\sqrt{s_{NN}}$ = 200 GeV and the pseudorapidity dependence of high p(T)
  suppression}},
\newblock Phys. Rev. Lett. \textbf{91}, 072305 (2003),
\newblock \doi{10.1103/PhysRevLett.91.072305}.

\bibitem{Arsene:2004ux}
I.~Arsene \emph{et~al.},
\newblock \emph{{On the evolution of the nuclear modification factors with
  rapidity and centrality in d+Au collisions at $\sqrt{s_{NN}}$ = 200 GeV}},
\newblock Phys. Rev. Lett. \textbf{93}, 242303 (2004),
\newblock \doi{10.1103/PhysRevLett.93.242303}.

\bibitem{Adler:2004eh}
S.~S. Adler \emph{et~al.},
\newblock \emph{{Nuclear modification factors for hadrons at forward and
  backward rapidities in d+Au collisions at $\sqrt{s_{NN}}$ = 200 GeV}},
\newblock Phys. Rev. Lett. \textbf{94}, 082302 (2005),
\newblock \doi{10.1103/PhysRevLett.94.082302}.

\bibitem{Adams:2006uz}
J.~Adams \emph{et~al.},
\newblock \emph{{Forward neutral pion production in p+p and d+Au collisions at
  $\sqrt{s_{NN}}$ = 200 GeV}},
\newblock Phys. Rev. Lett. \textbf{97}, 152302 (2006),
\newblock \doi{10.1103/PhysRevLett.97.152302}.

\bibitem{Adare:2011sc}
A.~Adare \emph{et~al.},
\newblock \emph{{Suppression of back-to-back hadron pairs at forward rapidity
  in $d+$Au collisions at $\sqrt{s_{NN}}$ = 200 GeV}},
\newblock Phys. Rev. Lett. \textbf{107}, 172301 (2011),
\newblock \doi{10.1103/PhysRevLett.107.172301}.

\bibitem{Gribov:1984tu}
L.~V. Gribov, E.~M. Levin and M.~G. Ryskin,
\newblock \emph{{Semihard Processes in QCD}},
\newblock Phys. Rept. \textbf{100}, 1 (1983),
\newblock \doi{10.1016/0370-1573(83)90022-4}.

\bibitem{Vitev:2007ve}
I.~Vitev,
\newblock \emph{{Non-Abelian energy loss in cold nuclear matter}},
\newblock Phys. Rev. C \textbf{75}, 064906 (2007),
\newblock \doi{10.1103/PhysRevC.75.064906}.

\bibitem{Kang:2011bp}
Z.-B. Kang, I.~Vitev and H.~Xing,
\newblock \emph{{Dihadron momentum imbalance and correlations in d+Au
  collisions}},
\newblock Phys. Rev. D \textbf{85}, 054024 (2012),
\newblock \doi{10.1103/PhysRevD.85.054024}.

\bibitem{Strikman:2010bg}
M.~Strikman and W.~Vogelsang,
\newblock \emph{{Multiple parton interactions and forward double pion
  production in pp and dA scattering}},
\newblock Phys. Rev. D \textbf{83}, 034029 (2011),
\newblock \doi{10.1103/PhysRevD.83.034029}.

\bibitem{McLerran:1993ni}
L.~D. McLerran and R.~Venugopalan,
\newblock \emph{{Computing quark and gluon distribution functions for very
  large nuclei}},
\newblock Phys. Rev. D \textbf{49}, 2233 (1994),
\newblock \doi{10.1103/PhysRevD.49.2233}.

\bibitem{McLerran:1993ka}
L.~D. McLerran and R.~Venugopalan,
\newblock \emph{{Gluon distribution functions for very large nuclei at small
  transverse momentum}},
\newblock Phys. Rev. D \textbf{49}, 3352 (1994),
\newblock \doi{10.1103/PhysRevD.49.3352}.

\bibitem{Guzey:2004zp}
V.~Guzey, M.~Strikman and W.~Vogelsang,
\newblock \emph{{Observations on dA scattering at forward rapidities}},
\newblock Phys. Lett. B \textbf{603}, 173 (2004),
\newblock \doi{10.1016/j.physletb.2004.10.033}.

\bibitem{Kharzeev:2004bw}
D.~Kharzeev, E.~Levin and L.~McLerran,
\newblock \emph{{Jet azimuthal correlations and parton saturation in the color
  glass condensate}},
\newblock Nucl. Phys. A \textbf{748}, 627 (2005),
\newblock \doi{10.1016/j.nuclphysa.2004.10.031}.

\bibitem{Marquet:2007vb}
C.~Marquet,
\newblock \emph{{Forward inclusive dijet production and azimuthal correlations
  in p(A) collisions}},
\newblock Nucl. Phys. A \textbf{796}, 41 (2007),
\newblock \doi{10.1016/j.nuclphysa.2007.09.001}.

\bibitem{STAR:2018iyz}
J.~Adam \emph{et~al.},
\newblock \emph{{Longitudinal Double-Spin Asymmetries for $\pi^{0}$s in the
  Forward Direction for 510 GeV Polarized $pp$ Collisions}},
\newblock Phys. Rev. D \textbf{98}(3), 032013 (2018),
\newblock \doi{10.1103/PhysRevD.98.032013}.

\bibitem{Whitten:2008zz}
C.~A. Whitten,
\newblock \emph{{The beam-beam counter: A local polarimeter at STAR}},
\newblock AIP Conf. Proc. \textbf{980}(1), 390 (2008),
\newblock \doi{10.1063/1.2888113}.

\end{thebibliography}

\nolinenumbers

\end{document}